\documentclass[12pt]{iopart}
\pdfoutput=1 
\usepackage{siunitx}
\usepackage{booktabs}
\usepackage{dcolumn}
\usepackage{bm}
\usepackage{multirow}

\usepackage{enumitem}

\usepackage{graphicx}

\usepackage{float}
\usepackage[nodisplayskipstretch]{setspace}
\usepackage{zref-perpage}
\zmakeperpage{footnote}
\usepackage[colorlinks,linkcolor=blue,citecolor=magenta]{hyperref}
\usepackage{subcaption}

\newcommand{\commentCWS}[1]%
{\textsf{\textcolor{blue}{#1$^{\mathrm{CWS}}$}}}
\newcommand{\commentJR}[1]%
{\textsf{\textcolor{red}{#1$^{\mathrm{JR}}$}}}

\begin{document}


\title{Carbon Diffusion in Concentrated Fe-C Glasses}
\author{Siavash Soltani$^{1}$, J\"org Rottler$^{2,3}$, Chad W.~Sinclair$^{1}$}
\address{$^{1}$Department of Materials Engineering, The University of British Columbia, Vancouver, BC, Canada V6T 1Z4,  \\
$^{2}$Department of Physics and Astronomy, The University of British Columbia, Vancouver, BC, Canada V6T 1Z1\\
$^{3}$Stewart Blusson Quantum Matter Institute, The University of British Columbia, Vancouver, BC, Canada V6T 1Z4
}

\ead{Chad.Sinclair@ubc.ca}

\begin{abstract}
    By combining atomistic simulations with a detailed analysis of individual atomic hops, we show that the diffusion of carbon in a binary Fe-C glass exhibits strong (anti-)correlations and is largely determined by the local environment. Higher local carbon concentrations lead to slower atomic mobility. Our results help explain the increasing stability of Fe-C (and other similar metal-metalloid glasses) against crystallization with increasing carbon concentration.
\end{abstract}

\section{Introduction}
Unlike most crystalline materials, the far from equilibrium nature of metallic glasses induces a significant time dependence to structural and functional properties.  Diffusion controlled de-vitrification, phase separation and structural aging all impact the ability to fabricate materials with technologically desirable properties that are stable under operating conditions. The fact that all technologically important metallic glasses are multicomponent alloys increases the complexity as one must consider the atomic mobility of all species and their interaction, under non-dilute conditions.  In the case of de-vitrification, solute repartitioning to or away from the growing crystalline phase can change the local composition,  atomic mobility and potentially the interfacial structure.  The rejection of metalloids like B and C from the growing crystalline phase in Fe-B and Fe-C glasses has been suggested to lead to a kinetic slowing of de-vitrification due to a strong dependence of diffusivity on the solute metalloid content in the glass \cite{Lawrence_2017}.    

The effect of solute `size' (e.g. metallic diameter) on atomic mobility in metallic glasses has been a point of discussion for some time.   Technologically the importance of this is driven by the key role metalloids (like C and B) play in all iron-based bulk metallic glasses \cite{suryanarayana2013iron}.  Metalloids not only improve glass forming ability \cite{nakajima1988crystallization, kanamaru1983preparation, bauer2004thermal}, but also increase the crystallization temperature \cite{nakajima1988crystallization, ruckman1980mossbauer,Lawrence_2017}.  Generally, though with several discrepancies, it is found from experiments that the activation enthalpy for diffusion is order of magnitude similar to that found for diffusion in crystals and increases with atomic size. The observation that, even for metalloids, the diffusivity is similar to that for crystalline materials led to early speculation that metalloid diffusion in glasses involved metal-metalloid exchange or coupled vacancy-metalloid motion. In the specific case of B diffusion in Fe$_{x}$B$_{1-x}$ glasses \cite{hasegawa1978iron,ruckman1980mossbauer}, it was suggested that B atoms occupy `interstitial-like' sites in low concentration and `substitutional-like' sites in high concentrations. As the substitutional-like character of the solute increases, more cooperative movement is required for diffusion of Fe and B, resulting in lower diffusivity and higher stability against crystallization with increasing solute concentration \cite{ruckman1980mossbauer}. This distinction between diffusion controlled by single atomic motion for `small' solute atoms (e.g.  \cite{geyer1995atomic}) and collective motion for `large' solute atoms (e.g. \cite{kluge2004diffusion,teichler2001structural,faupel1990pressure}) has now been more widely reported, though much more evidence exists for diffusion in systems with `large' solutes than `small'.

More recently, understanding the underlying correlation between composition and short to medium range ordering in metallic glasses has been facilitated by simulation (see e.g. \cite{sheng2012Relating,sheng2006atomic,ganesh2008ab}). Rather than a simple random packing of hard spheres, metallic glass alloys, and in particular metal-metalloid glasses, show a complex local structure which depends on composition and atom type. While atomistic simulations have shown that pure or dilute glasses are dominated by icosahedral short ranged order, addition of metalloid elements reduces the degree of icosahedral ordering, and the ordering of solvent atoms around the metalloid element takes very specific forms \cite{ganesh2008ab, sheng2006atomic}.  In the case of Fe based metallic glasses, the surroundings of the metalloid become associated with structures characteristic of the compositionally nearest intermetallic phase, e.g. tricapped trigonal prisms in Fe-B and Fe-C glasses. With increasing metalloid content the competition between the formation of metalloid centered structures and the icosehedral ordering preferred by the Fe leads to a form of  `geometrical frustration' \cite{ganesh2008ab}. The change in the local potential energy landscape associated with this change in structure impacts the atomic mobility and thus the bulk diffusivity.

In this work, we study the influence of carbon concentration on the carbon tracer diffusivity in model Fe-C glasses with molecular dynamics simulations. Going beyond a conventional analysis of mean-square displacements, we resolve individual C trajectories in terms of atomic jumps, which provides access to statistical distributions of jump lengths and times. The choice of the Fe-C system is motivated by our recent experimental work, which has shown strong effects of local carbon concentration on the crystallization kinetics in binary Fe-C glasses \cite{Lawrence_2017,fillon2015}.  Previously, the observed stasis in isothermal crystallization kinetics was attributed to thermodynamic effects \cite{fillon2015}. Here, we provide evidence that the slowing of diffusion due to the accumulation of carbon at the crystal/glass interface may be dominant.

\section{Computational Methods}

Molecular dynamics simulations were performed using LAMMPS \cite{plimpton1995fast} and the 2NN-MEAM Fe-C interatomic potential developed by Liyanage \emph{et al.} \cite{liyanage2014structural}.  This potential, built starting from the single element Fe and C potentials prepared by Lee \emph{et al.} \cite{lee2012atomistic} and Uddin \emph{et al.} \cite{uddin2010modified}, was explicitly fit to the properties of Fe$_3$C, including the behaviour of liquids containing up to 25~at.\%C.  All simulations reported here were performed under isothermal-isobaric (NPT) conditions employing a Nos\'e-Hoover thermotat/barostat.  A timestep of 2 fs was used throughout.

Glasses with three different carbon concentrations were generated starting from a periodic simulation box with dimensions of $\SI{51}{\angstrom} \times  \SI{67}{\angstrom} \times \SI{32}{\angstrom} $.  In all three cases, the starting structure was stoichiometric cementite, ${\rm Fe}_3{\rm C}$ (25 at.\%C), containing 8400 Fe atoms.  To obtain boxes containing 8 at.\% and 16 at.\% carbon, carbon atoms were randomly removed from the stoichiometric phase and a conjugate gradient minimization of the energy was performed to create the three starting materials.

\begin{figure}[htbp]
\centering
\includegraphics[width=0.5\textwidth]{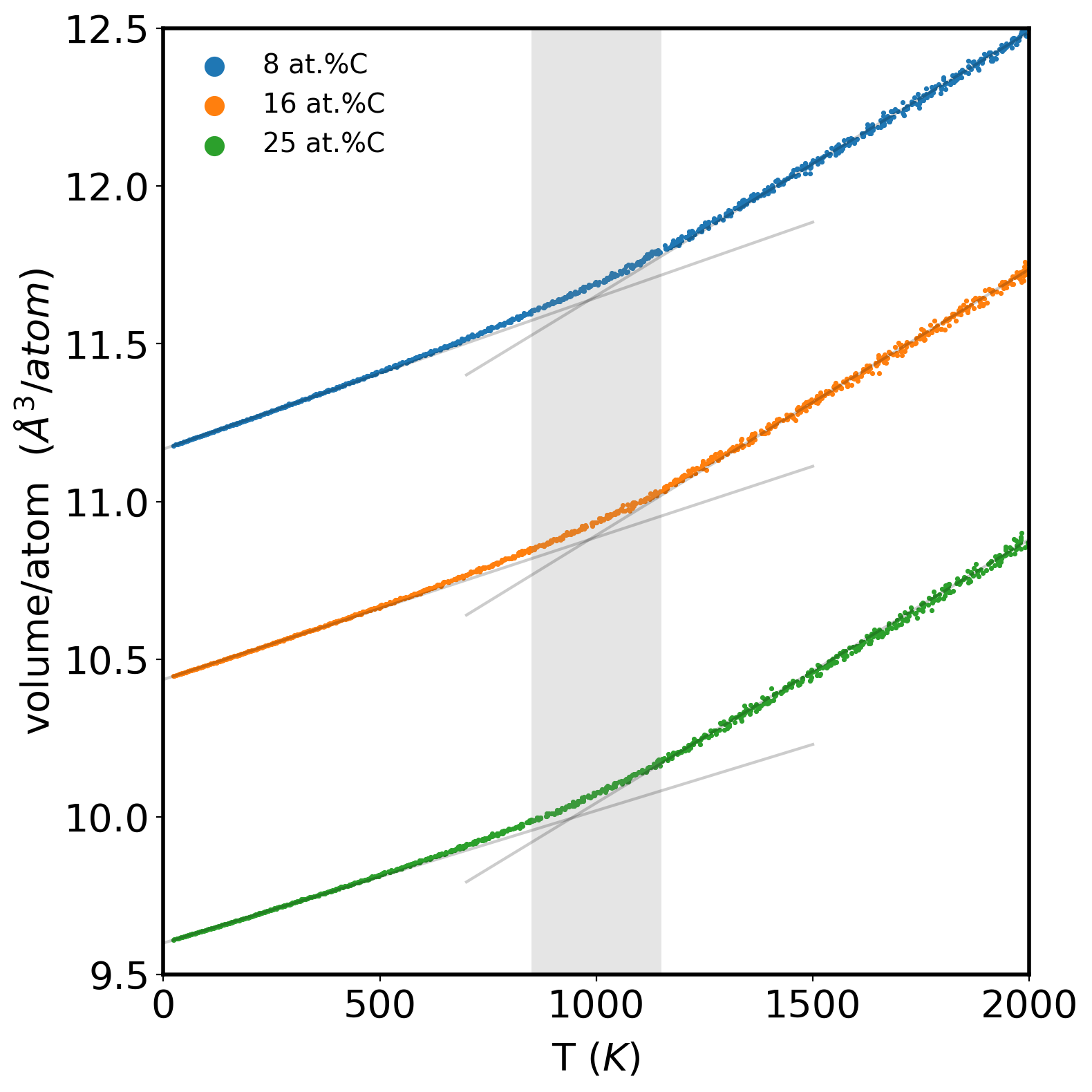}
\caption{Volume/atom as a function of temperature during quenching of the three alloys. The vertical grey line shows the temperature range 900 - 1100~K containing the glass transition temperature for all three compositions. }
\label{fig:Tg}
\end{figure}

The resulting models were melted by heating to T= 2000~K and held for 5~ns.  Following this, constant rate quenches were performed to below 100~K to form a glass (figure \ref{fig:Tg}). The slowest quench rate allowing all three compositions to be quenched without crystallization was found to be 600~K/ns.

The highest temperature at which data could be collected without crystallization of the most unstable (8 at.\%C) glass was 700~K.  All results shown here are therefore limited to 700~K.  Each run was heated to 700~K and held for 5~ns to equilibrate.  This was followed by a 200~ns hold during which the positions of all atoms were recorded every $\delta t =$ 0.1~ns.  The mean square displacement of carbon and iron atoms were calculated, treating the atoms as tracers, according to a time averaging scheme,
 
\begin{eqnarray}
\label{eq:MSD_Eq}
\left<( \Delta {\bf r}^{(i)} \left(t\right))^2\right> &=& \langle \left[{\bf r}^{(i)}\left(t_n\right) - {\bf r}^{(i)}\left(0\right)\right]^2 \rangle \\\nonumber
&=&  \frac{1}{N}\frac{\sum_{i=1}^N\sum_{j=1}^{n_{obs}-n}\left[{\bf r}^{(i)}\left(\left(n+j\right)\delta t\right) - {\bf r}^{(i)}\left(j\delta t\right)\right]^2}{n_{obs}-n}
\end{eqnarray}
We denote by ${\bf r}^{(i)}\left(t_n\right)$ the position of the $i^{th}$ atom at time $t_n = n \delta t$.  For a given atomic trajectory, $n_{obs}$ is the total number of observation times (i.e. 200 ns/0.1 ns for all cases here) and $N$ is the total number of observed (carbon or iron) atoms in the system.  A `hop detection algorithm' adopted from the work of Smessaert and Rottler \cite{smessaert2013distribution} was then applied to the same trajectories to resolve individual carbon atom `hops' or 'jumps', from which statistics on residence times, jump distances and jump correlations were studied. 
Ovito \cite{stukowski2009visualization} has been used for visualization as well as for the structural characterization of the glasses described below. 

\section{Results and Discussion}
\subsection{Structural characterization of the quenched glasses}
\begin{figure}[htbp]
\centering
\includegraphics[width=0.5\textwidth]{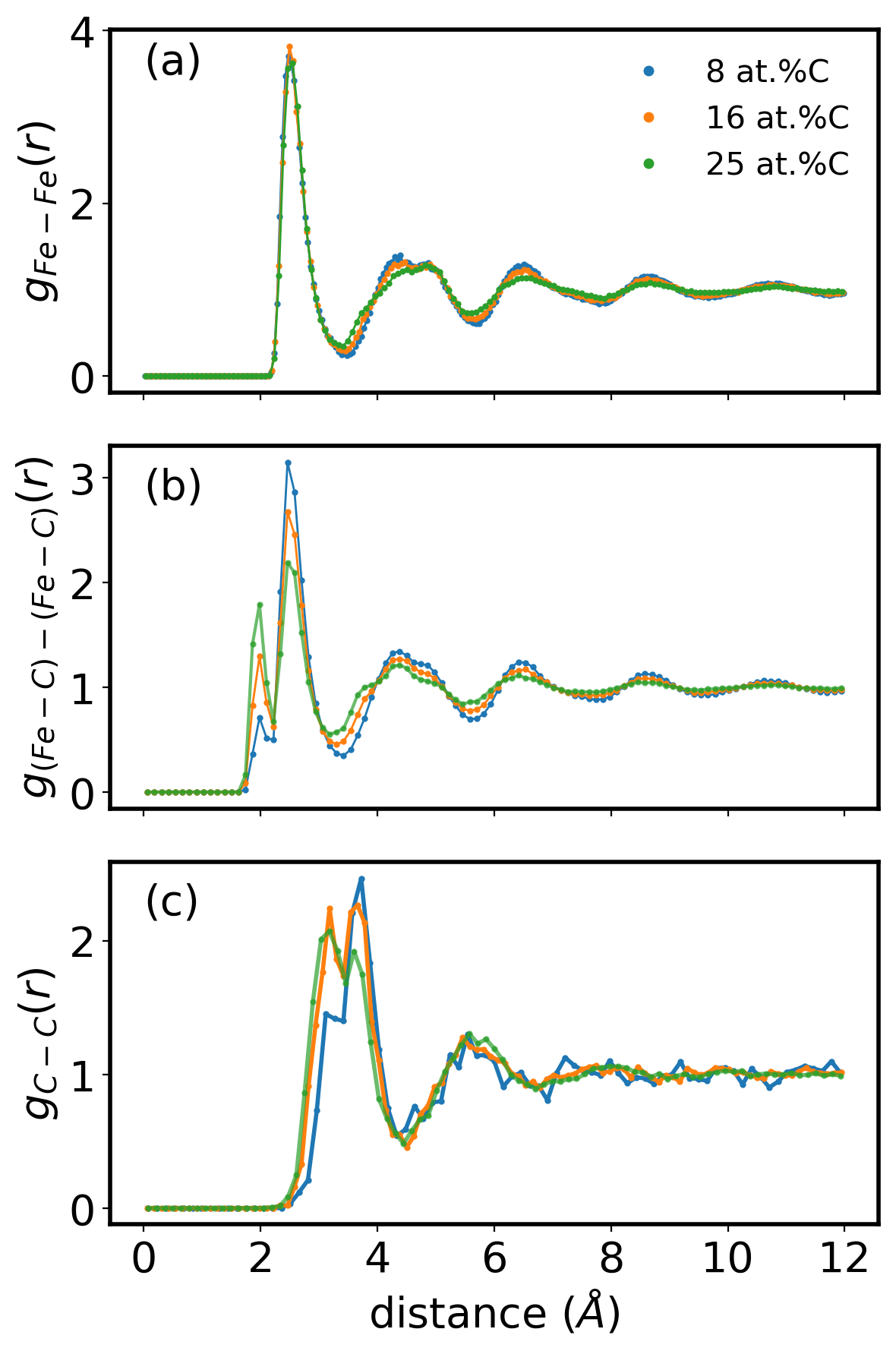}
\caption{ a) The Fe-Fe and b) the Fe-C and c) C-C partial radial distribution functions measured on the three quenched glasses. 
}
\label{fig:rdf}
\end{figure}

The three quenched glasses were found to exhibit many similarities. First, all three were found to have glass transition temperatures between $T_g \approx 900-1100~K$ (figure \ref{fig:Tg}).  No clear evidence of a composition dependence of $T_g$ could be observed.  The short to medium range atomic order of the glasses, reflected in the Fe-Fe, Fe-C and C-C partial radial distribution functions (figure \ref{fig:rdf}), also shows only minor differences.  In all three cases, the Fe-Fe nearest neighbour distance is \SI{2.6}{\angstrom} with a second doublet peak at \SI{4.7}{\angstrom}, this being similar to the Fe-Fe partial radial distribution function previously reported for Fe-B \cite{ganesh2008ab} and Ni-P \cite{sheng2006atomic} glasses as well as in Fe-C liquids \cite{pan2015atomic}.  The C-C partial radial distribution function shows a slightly larger nearest neighbour distance ($\sim$ 3.7 \AA) with a broader peak compared to that for Fe-Fe neighbours. Again, this is qualitatively similar to the results observed from simulations of Fe-B glasses \cite{ganesh2008ab} and Fe-C liquids \cite{pan2015atomic}. The only composition dependence can be found in the Fe-C partial radial distribution function.  Here, one sees a strengthening of the first peak in the rdf with increasing carbon content and a weakening of the second peak.   

The potential for longer ranged chemical heterogeneity, i.e. clustering or long-ranged ordering of carbon, was tested by calculating the number of carbon atoms within spherical volumes of radius $ r = \SI{4}{\angstrom}  $, centered on each carbon atom. A radius of $ r = \SI{4}{\angstrom}  $ was selected as it is located between the first and second nearest neighbour positions in the C-C partial radial distribution function (figure \ref{fig:rdf}) and thus identify carbon atoms within the nearest-neighbour coordination shell.  

\begin{figure}[htbp]
\centering
\includegraphics[width=0.5\textwidth]{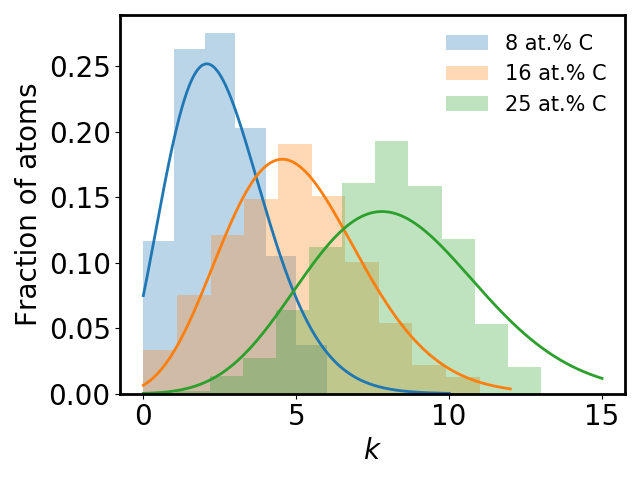}
\caption{Distribution of number of carbon neighbors of carbon atoms within a radial distance of $r = \SI{4}{\angstrom} $ ($k$). Solid lines show equation~(\ref{eqn:poisson}) where the values of $\bar{k}$ for each glass were calculated from the bulk compositions as, $\bar{k}\left(8\%C\right)= 2.6$, $\bar{k}\left(16\%C\right)= 5.1$ and $\bar{k}\left(25\%C\right)= 8.3$. }
\label{fig:poisson}
\end{figure}

If randomly distributed in space, the statistical distribution of carbon neighbours within the sampling volume should be Poissonian,
\begin{equation}
    P\left(k\right) = \frac{\bar{k}^{k}\exp{\left(-\bar{k}\right)}}{k!}
\label{eqn:poisson}
\end{equation}
where $k$ is the number of carbon atoms within an observation volume and $\bar{k}$ is the average expected observation number (average carbon concentration multiplied by the spherical observation volume). As shown in figure \ref{fig:poisson}, the agreement between equation~(\ref{eqn:poisson}) and the collected data is excellent, with only the sample containing 25 at.\%C showing deviations at the high number of neighbours. This likely reflects the repulsive nature of carbon atoms at close proximity.

\begin{figure}[htbp]
\centering
\includegraphics[width=0.9\textwidth]{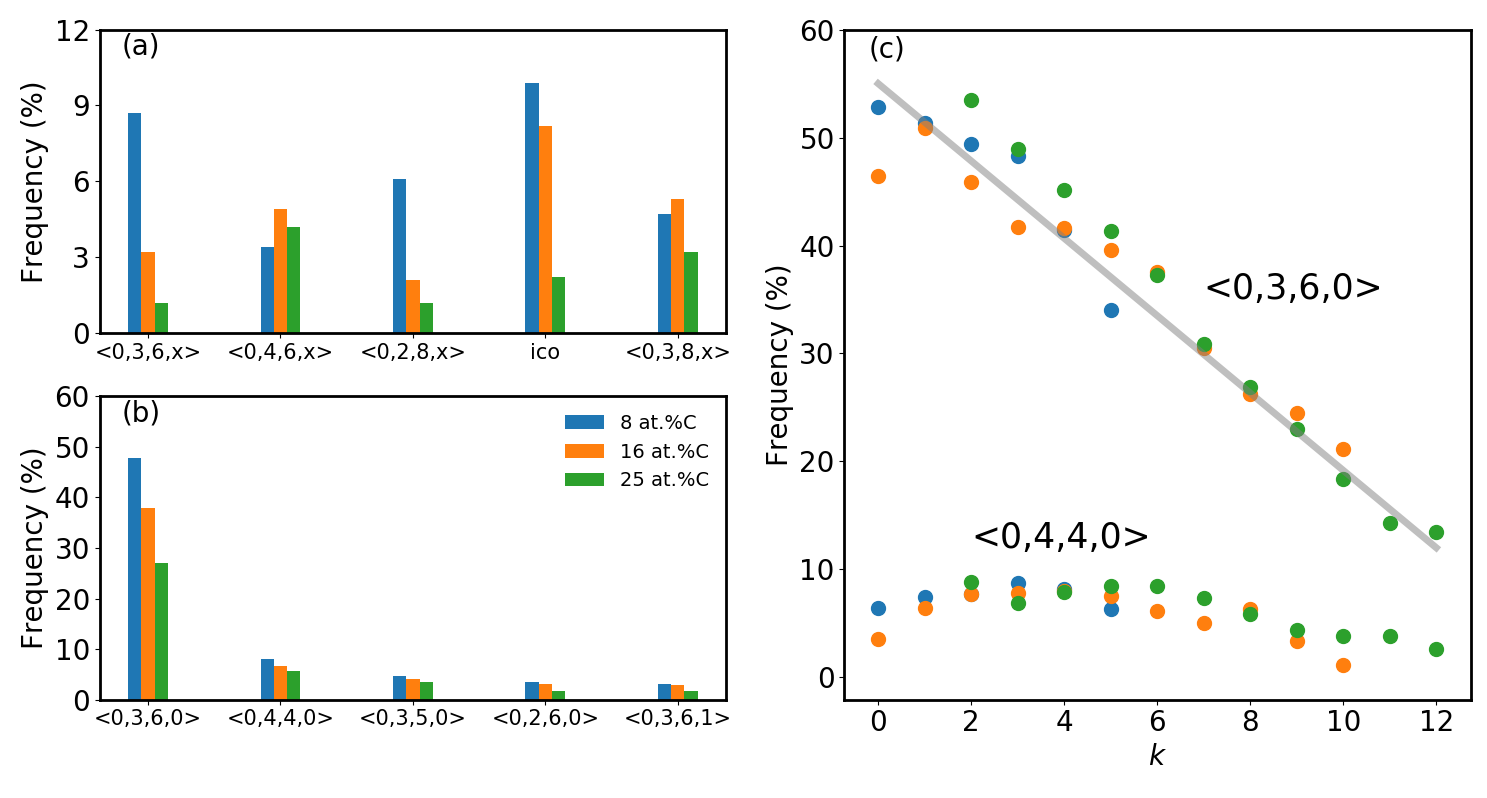}
\caption{Distribution of the most prevalent coordination polyhedra centered by (a) Fe atoms and (b) C atoms.  Statistics were generated by sampling structures every 10~ns, performing energy minimization followed by Voronoi calculations \cite{stukowski2009visualization}. Edges smaller than $\SI{0.3}{\angstrom} $ and faces smaller than $\SI{0.3}{\angstrom} ^{2}$ have been filtered out.  Panel (c) shows the frequency of $<0,3,6,0>$ and $<0,4,4,0>$ centered by C atoms as a function of number of carbon neighbors, $k$, within radial distance of $r = \SI{4}{\angstrom} $.}
\label{fig:voro}
\end{figure}

The carbon concentration dependence of the topological short-range order can be investigated in more detail by considering the coordination polyhedra, computed by Voronoi spatial tessellation,  centered on Fe and C atoms (figure \ref{fig:voro}). Unsurprisingly, the most common environment for Fe centred polyhedra was found to be icosahedral ($\left<0,0,12,0\right>$).  For the sample containing 8 at.\%C, the next most prevalent polyhedra belong to the icosahedral-like $\left<0,3,6,x\right>$ and $\left<0,2,8,x\right>$ classes (where $x =0,1,2,3,4$). With increasing carbon content, the fraction of these icosahedral (and icosahedral-like) structures decreases, this being consistent with the `geometrical frustration' observed for increasing metalloid content in other metal-metalloid glasses (see e.g. \cite{ganesh2008ab}).  

In the case of C centred polyhedra, by far the most commonly encountered environment was $\left<0,3,6,0\right>$, figure \ref{fig:voro}(b).  This represents a tri-capped trigonal prism arrangement around the carbon atom, which is the basic structural element forming the cementite (Fe$_3$C) structure.  The prevalence of this ordering has been previously noted in simulations of Fe-C liquids \cite{pan2015atomic} and in the case of B centered environments in Fe-B glasses \cite{ganesh2008ab}.  The prevalence of the $\left<0,3,6,0\right>$ environment drops rapidly with increasing carbon content of the glass, this also being a strong indicator of the increasing geometrical frustration experienced in the glass.

The geometrical frustration noted above, can be seen in a more local fashion if rather than correlating the polyhedra type with bulk composition we correlate it with the local composition.  In figure \ref{fig:voro} (c) we show the prevelance of the two most common carbon-centered polyhedra ($\left<0,3,6,0\right>$ and $\left<0,4,4,0\right>$) as a function of the number of carbon atoms within a radial distance of $ r = \SI{4}{\angstrom}  $.  The analysis performed for the three different glasses shows that, universally, the fraction of $\left<0,3,6,0\right>$ centered polyhedra decreases with local carbon fraction.  Interestingly, this suggests that the change in the local topological arrangement around a carbon atom is a unique function of the number of nearest neighbour carbon atoms rather than bulk composition.  The change in fraction of $\left<0,3,6,0\right>$ polyhedra as a function of the bulk carbon content (figure \ref{fig:voro} (b)) is therefore a reflection of the proportion of number of carbon-carbon neighbours (cf. figure \ref{fig:poisson}) rather than a reflection of some larger scale change in the structure of the glass. 

\subsection{Atomic Mobility: Mean Square Displacement}
\begin{figure}[htbp]
\centering
\includegraphics[width=0.5\textwidth]{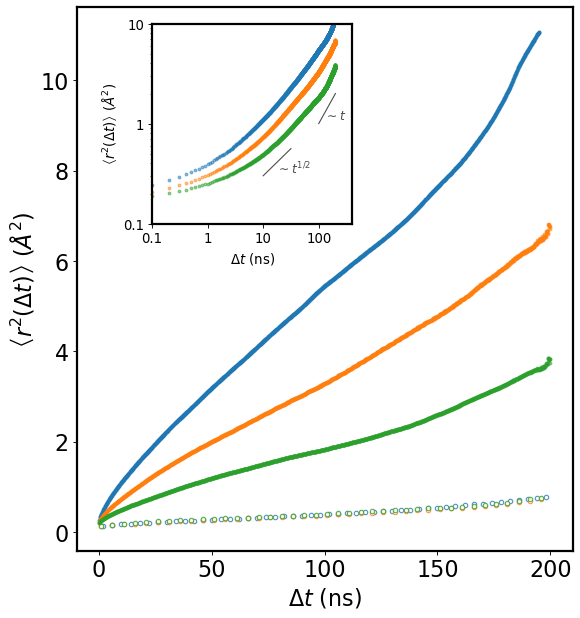}
\caption{Mean square displacement (MSD) of atoms in systems containing 8~at.\%C (green), 16~at.\%C (orange) and 25~at.\%C (blue). Solid points correspond to the MSD of carbon atoms while the open symbols correspond to the MSD of iron atoms.  The inset shows the mean square displacement of carbon atoms on a log-log scale showing the approach to linear (normal diffusive) scaling at times $>$ 100 ns.  }
\label{fig:MSD}
\end{figure}

During the 700~K isothermal-isobaric holds, the atomic trajectories for both Fe and C atoms were used to compute mean square displacements (eq.~(\ref{eq:MSD_Eq})).  The results are shown in figure \ref{fig:MSD}.  In the case of the Fe atoms, the mean square displacement remains small ($<$ \SI{1}{\angstrom^2} ) and sub-linear with respect to time over times of up to 200~ns.  No dependence of the mean square displacement on the carbon content was observed in this case. 

By contrast, the mean square displacement of carbon atoms exhibits a strong dependence of the  carbon concentration of the alloy.  In all three cases, a significant regime of sub-diffusive ($t \sim t^\alpha$, $\alpha < 1$) behaviour is observed with linear, diffusive scaling, being approached asymptotically at the longest observation times (see the inset to figure \ref{fig:MSD}).  Taking a simple linear fit to the mean square displacement over the last 100~ns of the trajectory indicates that the diffusivity of the 16~at.\%C sample is 60\% that of the 8~at.\%C sample, while the 25~at.\%C is 35\% that of the 8~at.\% sample. In accord with the discussion in the introduction, the mean square displacements observed here are within the same order of magnitude as those expected for carbon diffusion in crystalline BCC $\alpha$-Fe. The concentration dependence of the mean square displacement is, however, smaller for the glass compared to the composition dependence of C diffusion in BCC $\alpha$-Fe at the same temperature.  In the case of diffusion in the crystal, long range elastic repulsion between carbon atoms is predicted to significantly increase the barrier for diffusion as the carbon concentration is raised.  Experimentally, however, the observation of this composition dependence is not encountered owing to the very limited solubility of carbon in $\alpha$-Fe.

\subsection{Trajectory decomposition}

Although figure \ref{fig:MSD} reveals the anticipated decrease in carbon mobility with increasing carbon concentration, it does not provide any insight into the origins of this dependence. To delve further into the reasons for these trends, we have identified individual diffusive carbon jumps.  From this it is possible to reconstruct each carbon atom's trajectory with knowledge of distributions of jump rates, jump distances and correlations between jumps. Such analysis has proven to be useful in elucidating various relaxation phenomena, for instance structural aging in strong \cite{vollmayr-lee2013aging} and fragile \cite{warren2013aging} glass formers.

\begin{figure}[htbp]
\centering
\includegraphics[scale=0.4]{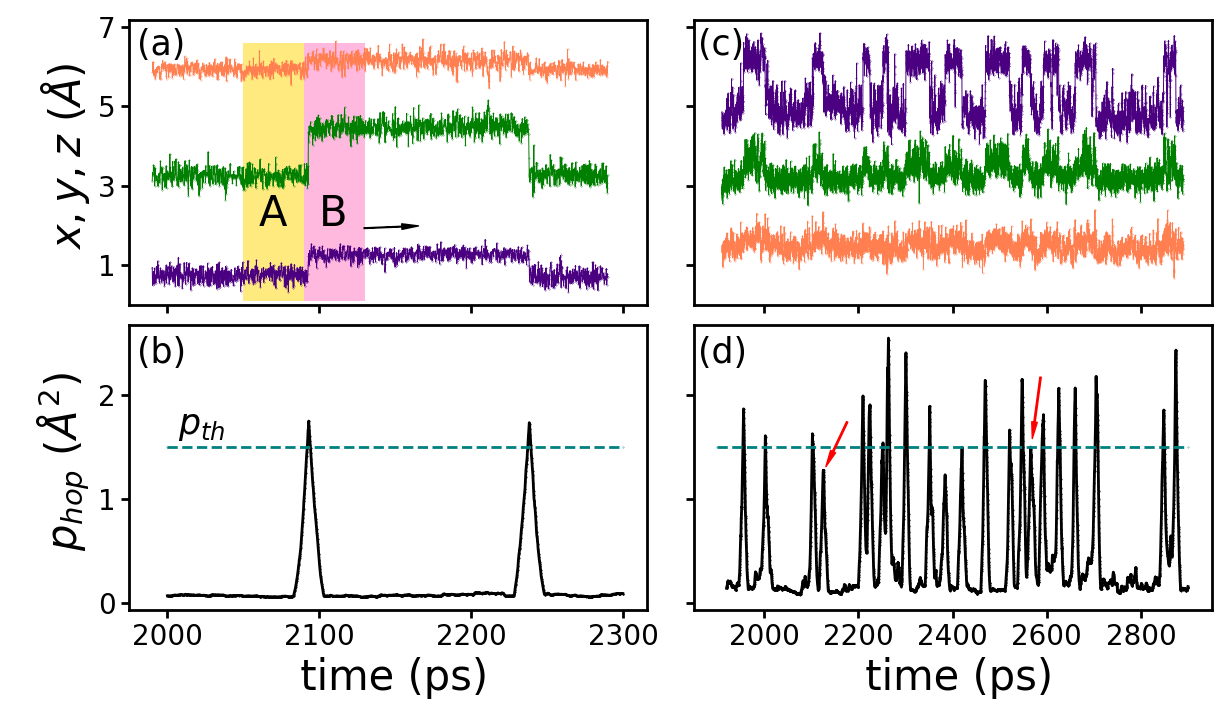}
\caption{ a) Partial trajectory of a carbon atom, the x,y and z coordinates of its position being shown. The evaluation time window and sections A and B are also highlighted (not to scale). The evaluation time window sweeps the entire trajectory. b) $p_{hop}$ profile of the trajectory shown in (a). The dashed lines show the threshold value $p_{th}$. c) Trajectory of an atom which makes several jumps to an adjacent cage and then quickly jumps back. The hop algorithm does not detect all backward jumps, as indicated with arrows in (d).}
\label{fig:p_hop}
\end{figure}

To capture the diffusive hops of C atoms during the isothermal annealing, the positions of C atoms were saved every $N_{obs}=100$ time steps, and a hop detection algorithm was implemented following ref.~\cite{smessaert2013distribution}. For each C atom, starting from $t=0$, a time window containing $N_{hist}= 100$ frames of the trajectory is input to the algorithm. The algorithm divides this evaluation time window into two sections $A[0,N_{hist}/2)$ and $B[N_{hist}/2,N_{hist})$, each containing $N_{hist}/2$ frames (see figure \ref{fig:p_hop}) and calculates a hop identifier parameter, $p_{hop}$ according to \cite{smessaert2013distribution},
\begin{equation}\label{eq:p_hop}
p_{hop}(t) = \sqrt{\langle (r_A-\overline{r}_B)^2 \rangle.\langle (r_B-\overline{r}_A)^2 \rangle  }
\end{equation}

which is a measure of average distance between the mean position $\overline{r}_A$ in section A, and all trajectory points in section B, $r_B$, and vice versa. The oldest time frame is then removed and the next time frame of the trajectory is added to the end of the evaluation time window to keep the number of frames in the evaluation time window $N_{hist}$ frames. The procedure is repeated until the final time frame of the trajectory.  The power of this algorithm is that $p_{hop}$ changes rapidly when a hop occurs, as shown in figure \ref{fig:p_hop} (b). A threshold $p_{th}$ is then defined and a hop is identified when $p_{hop} > p_{th}$. The time of the jump is then recorded where $p_{hop}$ is at its maximum and the jump vector, ${\bf l}= \overline{r}_B-\overline{r}_A$ can be obtained using the mean positions in section A and B. It should be mentioned that $N_{hist}$ sets the resolution of the algorithm. A large value results in low resolution, whereas a small value causes small fluctuations in displacement to produce large peaks in the profile of $p_{hop}$.

Frequently, atoms were observed to jump from one cage to another only to return quickly to their starting point. An example of such behaviour is shown in figure \ref{fig:p_hop} (c) and (d). Since the size of the evaluation time window, $N_{hist}$, sets the resolution of the algorithm, a hop with smaller residence time than sections A or B ($N_{hist}/2$), can result in a $p_{hop} < p_{th}$. Reducing $N_{hist}$ can allow one to capture these short lived jumps, though at significant computational cost. To ensure all hops are recorded one can also reduce $p_{th}$.  However, as one does this the distinction between a cage breaking jump and position fluctuations within a cage breaks down. To overcome these problems, a value of $p_{th}= \SI{1.5}{\angstrom} ^{2}$ was fixed which easily captures hops with long sojourn time (figure \ref{fig:p_hop}(b)). This value can however lead to missing some quick backward jumps as shown in figure \ref{fig:p_hop}(d). Therefore the position of each atom obtained from the hop algorithm and the actual position from MD snapshots were compared every 20~ns. If these two positions differed by more than \SI{2}{\angstrom}, all jumps recorded during that time period were considered to be non-diffusive, rapid back and forth hops, and subsequently deleted from the recorded jumps. In this way, we ensure that we properly track the trajectory of each atom while removing short lived `flickers' which do not contribute to diffusion from the list of events.

\begin{figure}[htbp]
\begin{subfigure}{.45\textwidth}
  \centering
  \includegraphics[width=.8\linewidth]{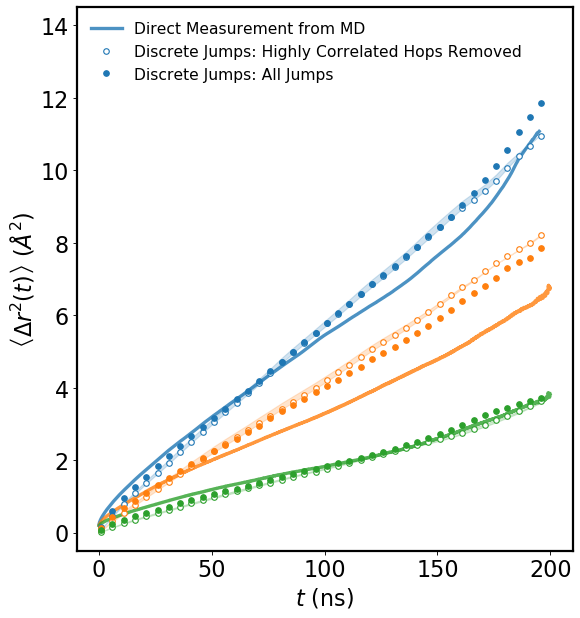}  
  \caption{}
  \label{fig:vanhove_msdplot}
\end{subfigure}
\begin{subfigure}{.45\textwidth}
  \centering
  \includegraphics[width=.85\linewidth]{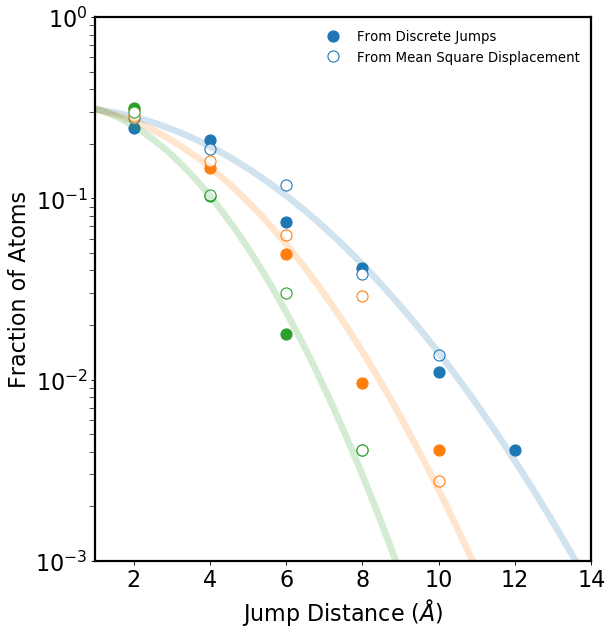} \caption{} 
  \label{fig:vanhoveplot}
\end{subfigure}
\caption{(a) Mean square displacement obtained directly from MD trajectory (solid line) compared to the MSD obtained from the hop detection algorithm. Open symbols show the results obtained when all detected hops are used (equation~(\ref{eqn:MSDdiscrete})) while closed symbols show results when strongly correlated jumps are removed from the trajectory (equation~(\ref{eqn:onlyuncorrhops})). For the results from uncorrelated jumps only the symbols represent the case where a threshold of $\cos{\left(\theta\right)} > -0.8$ was used.  The shaded band shows the variation if values of -0.7 or -0.9 are used instead. (b) The distribution of atomic displacements (van Hove distribution) after $t=\SI{200}{ns}$. Solid circles are obtained directly from the atom positions at time = 0 and time = 200~ns.  The open symbols are the displacements obtained by summing the displacements obtained from the hop detection algorithm. The lines are best fit Gaussian distributions. Blue corresponds to 8~at.\%C, orange to 16~at.\%C and green to 25~at.\%C.}
\label{fig:vanhove}
\end{figure}

In order to check that the hop detection method indeed captures all relevant aspects of the diffusional motion, the mean square displacement was reconstructed from the detected jumps. To do this, the displacement of an atom $(i)$ having $N_i\left(t\right)$ discrete jumps, $j$, each having a jump vector $\mathbf{l}_j^{(i)}(t)$ is computed as
$\Delta\mathbf{r}^{\left(i\right)}\left(t\right) = \sum_j^{N_i\left(t\right)}\mathbf{l}_j^{\left(i\right)}$.  With this assumption, the mean square displacement is

\begin{eqnarray}\label{eqn:MSDdiscrete}
     \left<(\Delta {\bf r}^{\left(i\right)}\left(t\right))^2\right> &=& 
     \left< \left(\sum_j^{N_i(t)}\mathbf{l}_j^{(i)}\right)
     \left(\sum_k^{N_i(t)}\mathbf{l}_k^{(i)}\right)\right>\\\nonumber
     &=&\left\langle\sum_j^{N_{i}(t)}({\bf l}^{(i)}_j)^2 \right\rangle + 2\left\langle\sum_j^{N_{i}(t)}\sum_{k>j}^{N_{i}(t)}{\bf l}^{(i)}_j{\bf l}^{(i)}_k\right\rangle
\end{eqnarray}

where the angle brackets $\left<\right> \equiv \left<\right>_i$ denote an average over all atoms as before.  Figure \ref{fig:vanhove_msdplot} shows the mean square displacement reconstructed from the detected atomic hops (open circles) alongside the data obtained directly from MD simulations (same data as shown in figure \ref{fig:MSD}), while figure \ref{fig:vanhoveplot} shows the distribution of displacements (Van Hove distribution) observed at the end of the annealing time, $t$=200~ns, obtained from both the atom positions at the beginning and end of the MD run and from the hop algorithm. The good agreement between the data obtained in these two ways provides confidence that the hop detection algorithm is able to adequately track the displacements of atoms.   

\begin{figure}[htbp]
\centering
\includegraphics[width=0.43\textwidth]{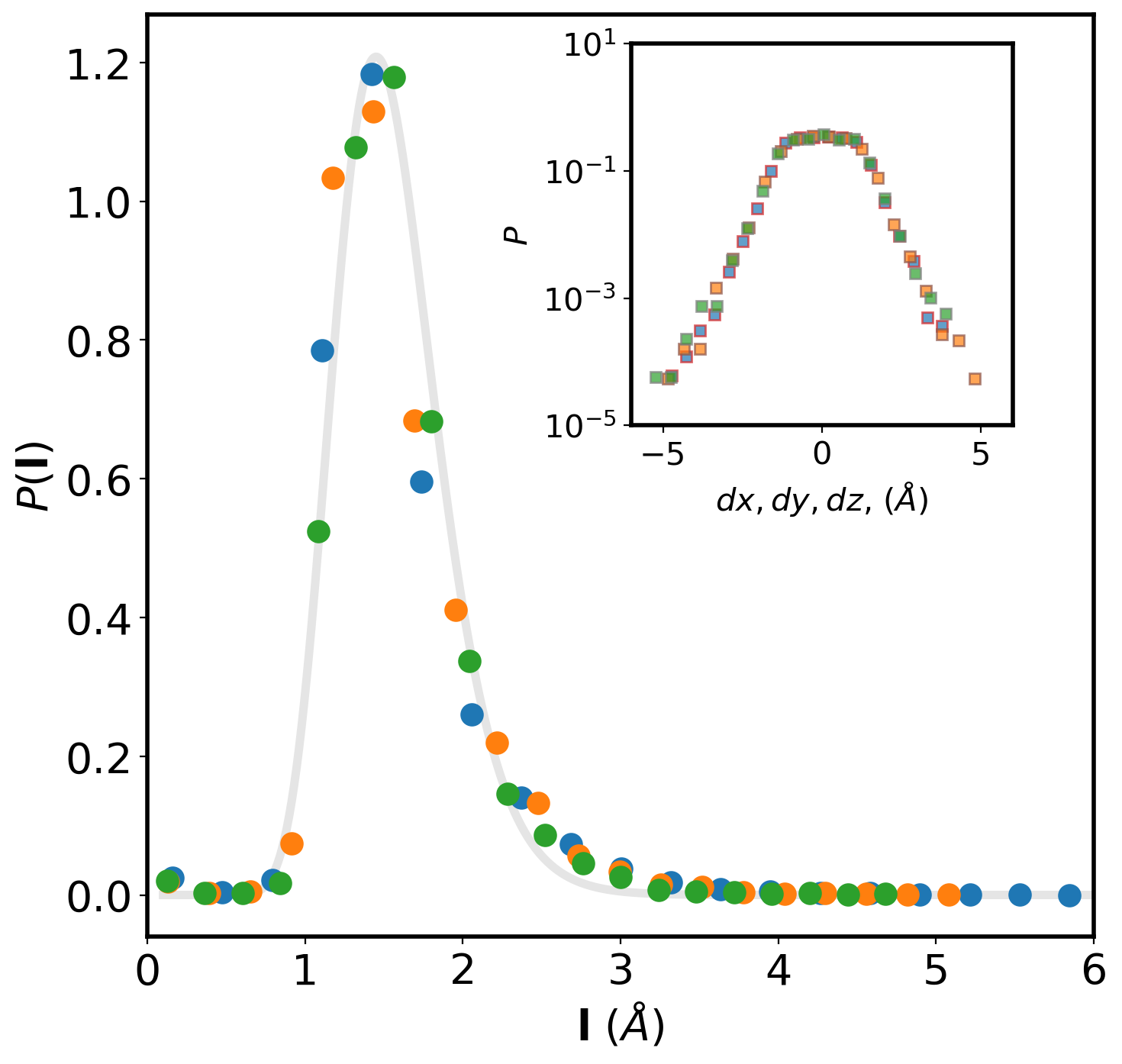}
\includegraphics[width=0.4\textwidth]{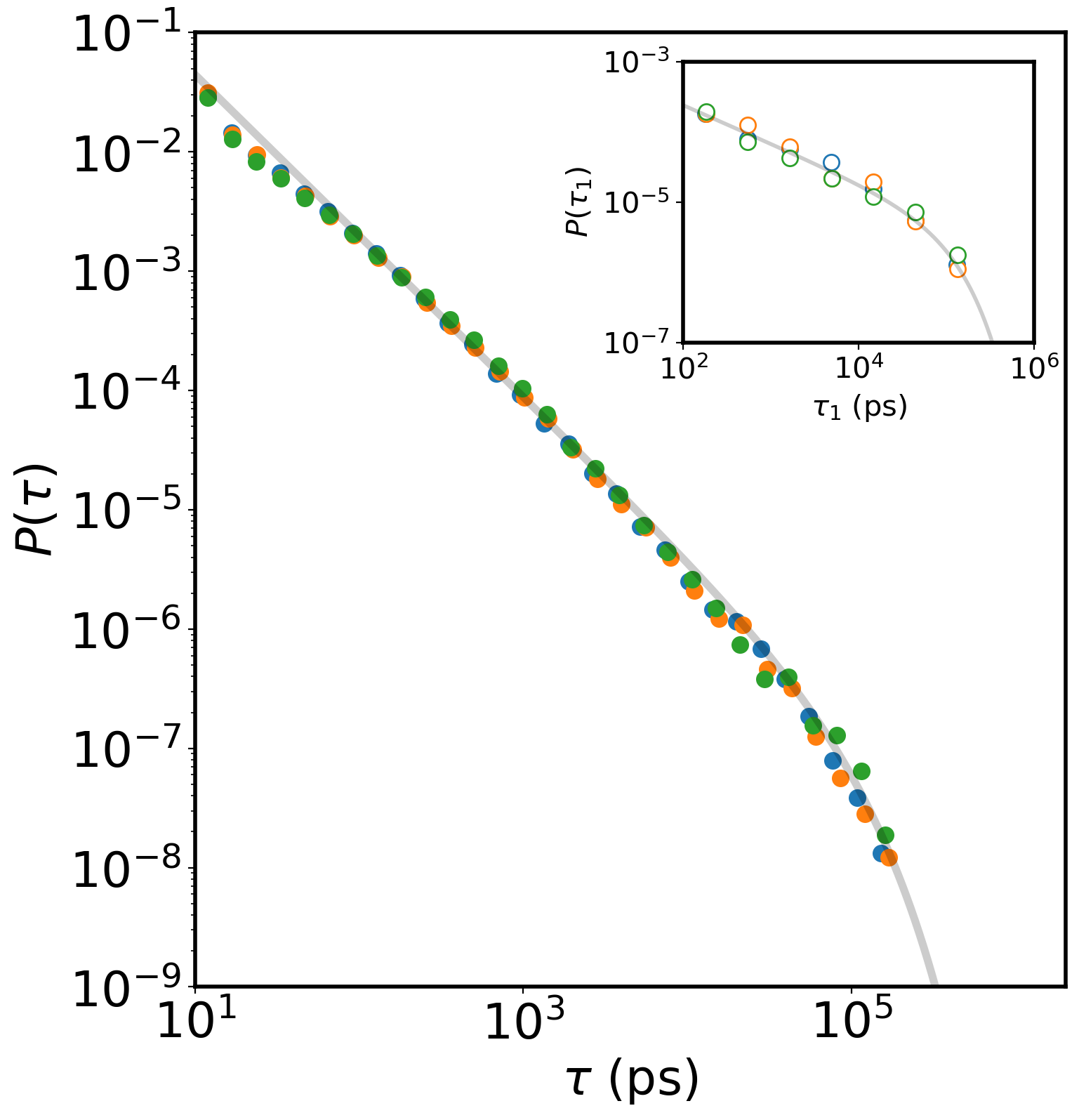}
\caption{(a) Distribution of jump lengths ($\ell =|{\bf l}|$) measured from the hop detection algorithm.  The average jump length is $\left<\ell\right> = \SI{1.58}{\angstrom} $. Blue corresponds to 8at.\%C, orange to 16at.\%C and green to 25 at.\%C.  The line is intended as a guide to the eye.  The inset shows the distribution of spatial components of the jump vectors for the sample containing 8 at.\%C sample. (b) Distribution residence times, $\tau$. The inset shows the distribution of first jump times ($\tau_1$). Blue corresponds to 8 at.\%C, orange to 16~at.\%C and green to 25 at.\%C.  }
\label{fig:jl_dx}
\end{figure}

From the hop detection algorithm it is possible to obtain separate statistics on the individual contributions to the mean square displacement arising from atomic jumps. Figure \ref{fig:jl_dx}(a) shows the distribution of the magnitudes of the jump lengths observed over the 200~ns anneal. It shows that the jump lengths for all three glasses fall within a narrow distribution from \SIrange[range-units=single]{1}{2}{\angstrom} with an average of \SI{1.58}{\angstrom}. The distribution of the components of the jump vector along the coordinate system of the simulation box are shown for the case of the sample containing 8 at.\%C in the inset. This distribution shows a plateau followed by an exponential decay, similar to that observed in other glasses \cite{helfferich2014continuous}. 


Figure \ref{fig:jl_dx}(b) shows the probability distribution function for the residence times, the distribution of times for the first observed jump shown in the inset, for the three glasses.  The distributions for all three cases collapse to the same (truncated) power law distribution (exponent between -1.3 and -1.4), commonly observed for a variety of glasses \cite{smessaert2013distribution, helfferich2014continuous, warren2009atomistic}.  While the probability distributions are nearly identical for the three glasses, the number of observed jumps per atom in the 200~ns simulations was not. For instance, the glass containing 8~at\%C produced nearly twice as many jumps/atom as the glass containing 25~at.\%C.

Many of these jumps are, however, often highly anticorrelated as already noted with respect to the partial trajectory shown in figure \ref{fig:p_hop}.  A jump in direction ${\bf l}$ is followed immediately by a reverse jump in direction $-{\bf l}$.  In order to characterize jumps for each particle as correlated or uncorrelated, $N_{i}(t)=N_{i,u}(t)+N_{i,c}(t)$, the following procedure was used.  For a given particle, two consecutive jumps were compared.  If the angle between the two consecutive jump vectors ($\theta_{i,i+1}$)  was greater than a critical angle then the two jumps were tagged as correlated.  The choice of this critical angle is arbitrary, though as shown below, the results are not drastically changed by its choice.  From the resulting list of uncorrelated jumps the atomic trajectory was reconstructed and a new set of jump vectors (${\bf l}^{(i)}_{j,u}$) calculated ensuring that the final position of each atom matched that obtained from the full set of jumps. If one considers only these uncorrelated hops, the cross terms in equation (\ref{eqn:MSDdiscrete}) vanish and we can approximate the mean-square displacement as
\begin{equation}
    \langle (\Delta {\bf r}^{(i)}(t))^2 \rangle = \langle\sum_j^{N_{i,u}(t)}({\bf l}^{(i)}_{j,u})^2 \rangle
\label{eqn:onlyuncorrhops}
\end{equation}
where the mean square jump length in this equation is obtained from the recalculated jumps as described above. The solid symbols in figure \ref{fig:vanhove_msdplot} show this approximation and track the open symbols from equation~(\ref{eqn:MSDdiscrete}) very well, which supports the decomposition. 

\begin{figure}[htbp]
\centering
\includegraphics[width=0.8\textwidth]{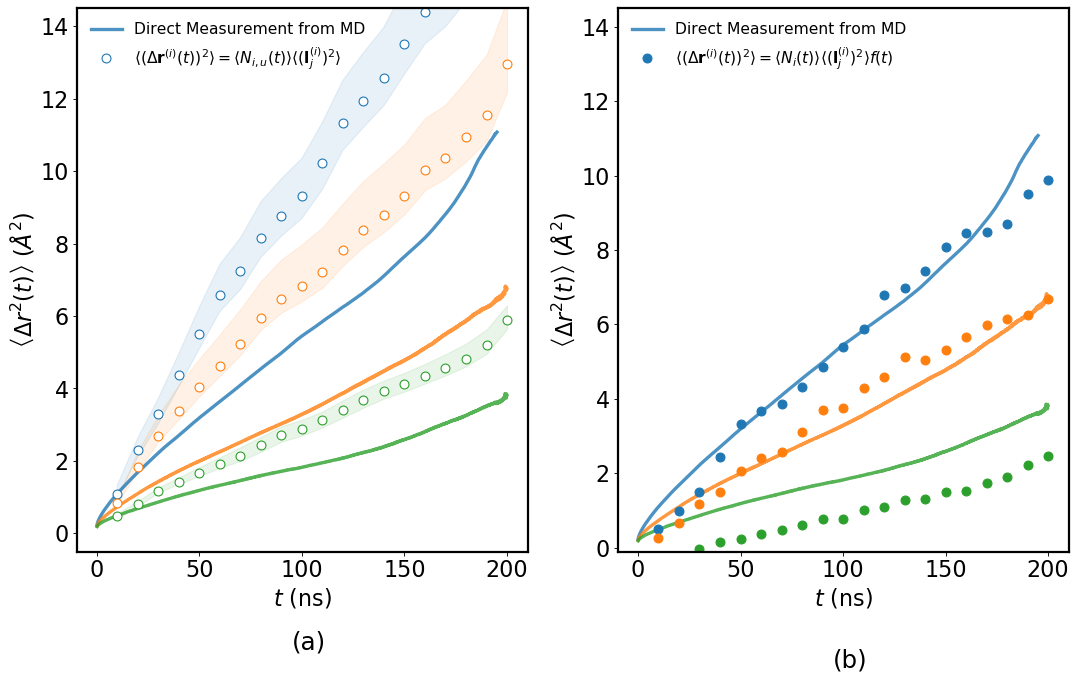}
\caption{a) The mean square displacement from direct MD observations (solid lines, the same results shown in figure \ref{fig:vanhove_msdplot}) compared to the results predicted using the average number of uncorrelated hops per atom, equation~(\ref{eqn:uncorr_factorized}).  The open symbols represent uncorrelated jumps identified as pairs of jumps where the angle between adjacent jumps fulfils the condition $\cos{\left(\theta_{i,i+1}\right)} > - 0.8$.  As this threshold is arbitrary, the results bounded by $\cos{\left(\theta_{i,i+1}\right)} > - 0.7$ and $ \cos{\left(\theta_{i,i+1}\right)} > -0.9$ are also shown as the shaded bands. b) The mean square displacement from direct MD observations (solid lines) compared to the predicted mean square displacement using the average total number of observed jumps per atom (correlated and uncorrelated) and the correlation factor, equation~(\ref{eqn:corrfac_factorized}). Blue corresponds to 8 at.\%C, orange to 16 at.\%C and green to 25 at.\%C. }
\label{fig:MSD_Nhops}
\end{figure}

Going one step further, we assume that on average all atoms make the same number of uncorrelated jumps (a strong assumption given the broad power law distribution of jump times) and that all of these jumps have the about the same squared jump length. The resulting expression,

\begin{equation}
    \langle (\Delta {\bf r}^{(i)}(t))^2 \rangle = \langle N_{i,u}(t)\rangle \langle ({\bf l}^{(i)}_{j,u})^2 \rangle.
    \label{eqn:uncorr_factorized}
\end{equation}
is evaluated in figure \ref{fig:MSD_Nhops}. The approximation still captures the overall trend with concentration, but systematically over predicts the mean-square displacement. It is notable that $\langle ({ l}^{(i)}_{j,u}) \rangle \approx 1.9$~\AA{}  was found to be significantly larger than $\langle ({ l}^{(i)}_{j}) \rangle \approx 1.6$~\AA{}  indicating that correlated jumps tend to be shorter than uncorrelated jumps.

In an alternative treatment of the correlated diffusive motion, one can explicitly separate correlated and uncorrelated jumps by rewriting equation~(\ref{eqn:MSDdiscrete}) as,

\begin{eqnarray}
     \left<\Delta {\bf r}^{\left(i\right)}\left(t\right)^2\right> 
     &=&\left<{\sum_j^{N_i\left(t\right)}\left(\mathbf{l}_j^{(i)}\right)^2}\right>\left(1+2\sum_n^{N_i\left(t\right)}C_{auto}(n)\right)
\label{eqn:MSDcorruncorr}
\end{eqnarray}
where
\begin{equation}
    C_{auto}(n) = \frac{\left< \sum_{j}^{N_i\left(t\right)-n}\mathbf{l}_j^{\left(i\right)}\mathbf{l}_{j+n}^{(i)}\right>}{\left<\sum_j^{N_i\left(t\right)}\left(\mathbf{l}_j^{\left(i\right)}\right)^2\right>}
\end{equation}
The first term on the right hand side of equation~(\ref{eqn:MSDcorruncorr}) is the uncorrelated mean square displacement; the mean square displacement that would be predicted if the observed jumps were selected in random order to construct the trajectory.  The second term contains all effects from correlations, $C_{auto}(n)$ being the jump auto-correlation function for jumps separated by $n$ jumps.  The ratio of the true MSD to the purely uncorrelated MSD is usually called the correlation factor
\begin{equation}
     f(t) =\langle (\Delta {\bf r}^{(i)}(t))^2 \rangle/ \langle\sum_j^{N_{i}(t)}({\bf l}^{(i)}_j)^2 \rangle
\label{eq:timecorrelation}
\end{equation}

\begin{figure}[htbp]
\centering
\includegraphics[width=0.9\textwidth]{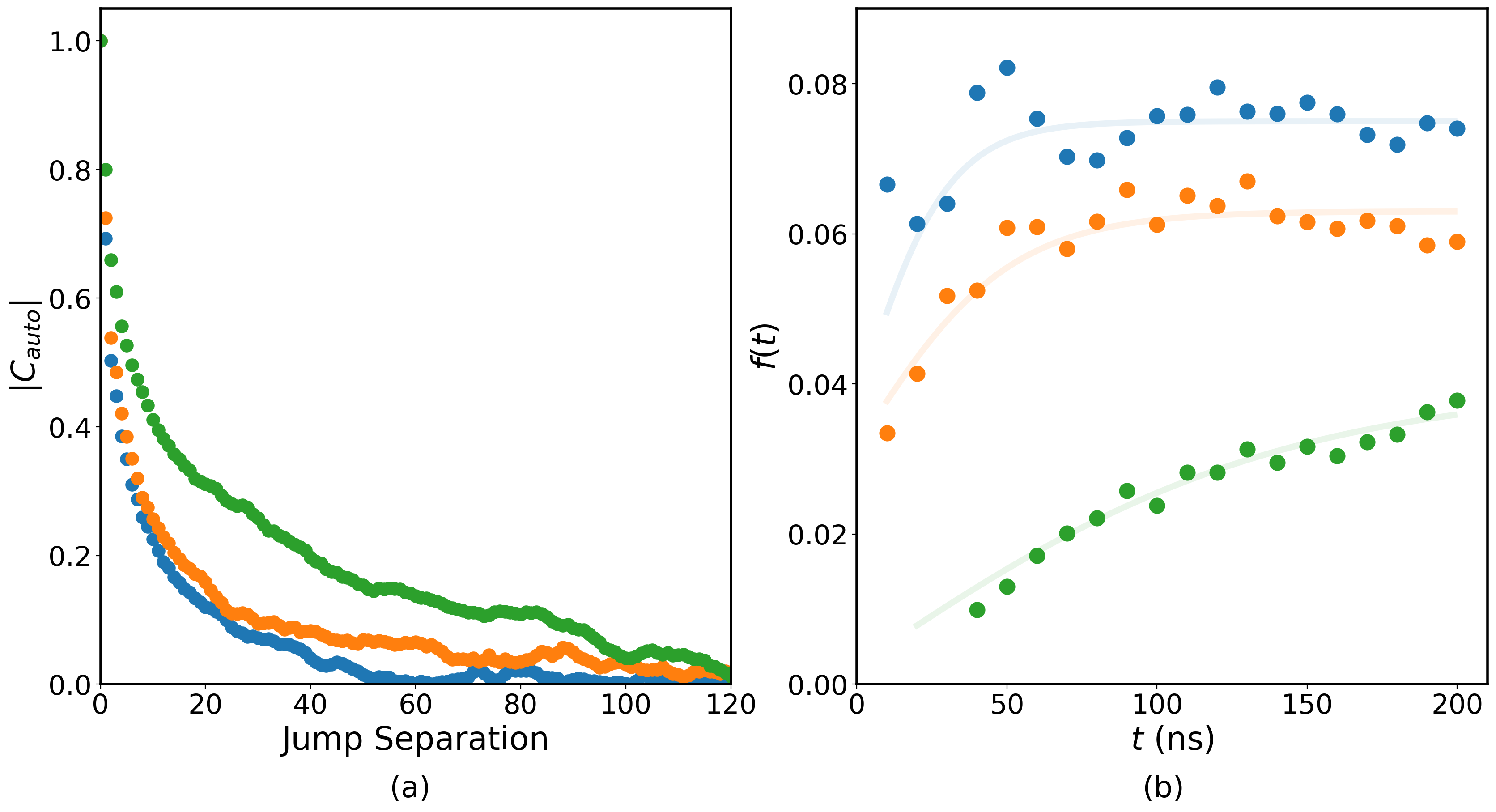}
\caption{a) Magnitude of the autocorrelation function for the three glasses studied. b) The time evolution of the correlation factor $f\left(t\right)$ evaluated according to equation~(\ref{eq:timecorrelation}). Here the lines are only intended as a guide to the eye.  Blue corresponds to 8~at.\%C, orange to 16~at.\%C and green to 25~at.\%C. }
\label{fig:corr}
\end{figure}

Figure \ref{fig:corr} shows the magnitude of the autocorrelation plotted as a function of the number of jumps separating two given jumps.  Here we see a significant difference between the three glasses.  The range of the correlation increases from 8~at.\%C to 25~at\%C.  For all three glasses the long correlation lengths reveal long periods of highly correlated jumps.   

It is also valuable to examine how $f(t)$ evolves with time. Figure \ref{fig:corr}b shows that for the glasses containing 8~at.\%C and 16~at.\%C the value of the correlation factor saturates within the first $\sim$50~ns of the simulation, while $f(t)$ for the glass containing 25~at.\%C still has still not reached a constant value by the end of the simulation. 

The picture that emerges is that carbon atoms make multiple coordinated jumps between two or more nearby `basins'.  Many such jumps tend to be made before the carbon atom will escape from this `super-basin'.  The highly correlated jumps made within one such `super-basin' then do not contribute to diffusive behaviour, this being reflected by the magnitude of the correlation factor.  Using the results in figure \ref{fig:corr} and equation (\ref{eqn:MSDcorruncorr}) we find correlation factors for the three glasses of $f_{8\%} = 0.06$, $f_{16\%}=0.04$ and $f_{25\%}=0.02$ after 200~ns.  

With values for the correlation factor in hand, we can propose yet another approximation of equation~(\ref{eqn:MSDdiscrete}), which consists again in assuming that on average all atoms jump about the same number of times and all these jumps have the about the same squared jump length. This yields
\begin{equation}
    \langle (\Delta {\bf r}^{(i)}(t))^2 \rangle = \langle N_{i}(t)\rangle \langle ({\bf l}^{(i)}_j)^2 \rangle f(t),
    \label{eqn:corrfac_factorized}
\end{equation}
the evaluation of which is shown in figure \ref{fig:MSD_Nhops}b, where $f(t)$ has been taken from the data shown in fig. \ref{fig:corr}b. This approximation captures the time and concentration dependence of the mean square displacement very well.  The only case not matching the direct molecular dynamics observations well being the 25~at.\%C glass, where it was already observed that the value of $f$ had not yet converged by the end of the run (fig. \ref{fig:corr}). 

The results so far indicate that the variation of the MSD with carbon concentration is carried by the number of jumps and the correlation factors. Our simulations allow to further resolve a dependence of the average number of jumps in terms of the local environment. As noted in the introduction, there has been an increasing recognition of the importance of the local structure of the glass on atomic mobility. Following the same logic as that used in the construction of figure \ref{fig:voro}c, we start by looking at the dependence of atomic motion on the local chemistry.

\begin{figure}[htbp]
\centering
\includegraphics[width=0.45\textwidth]{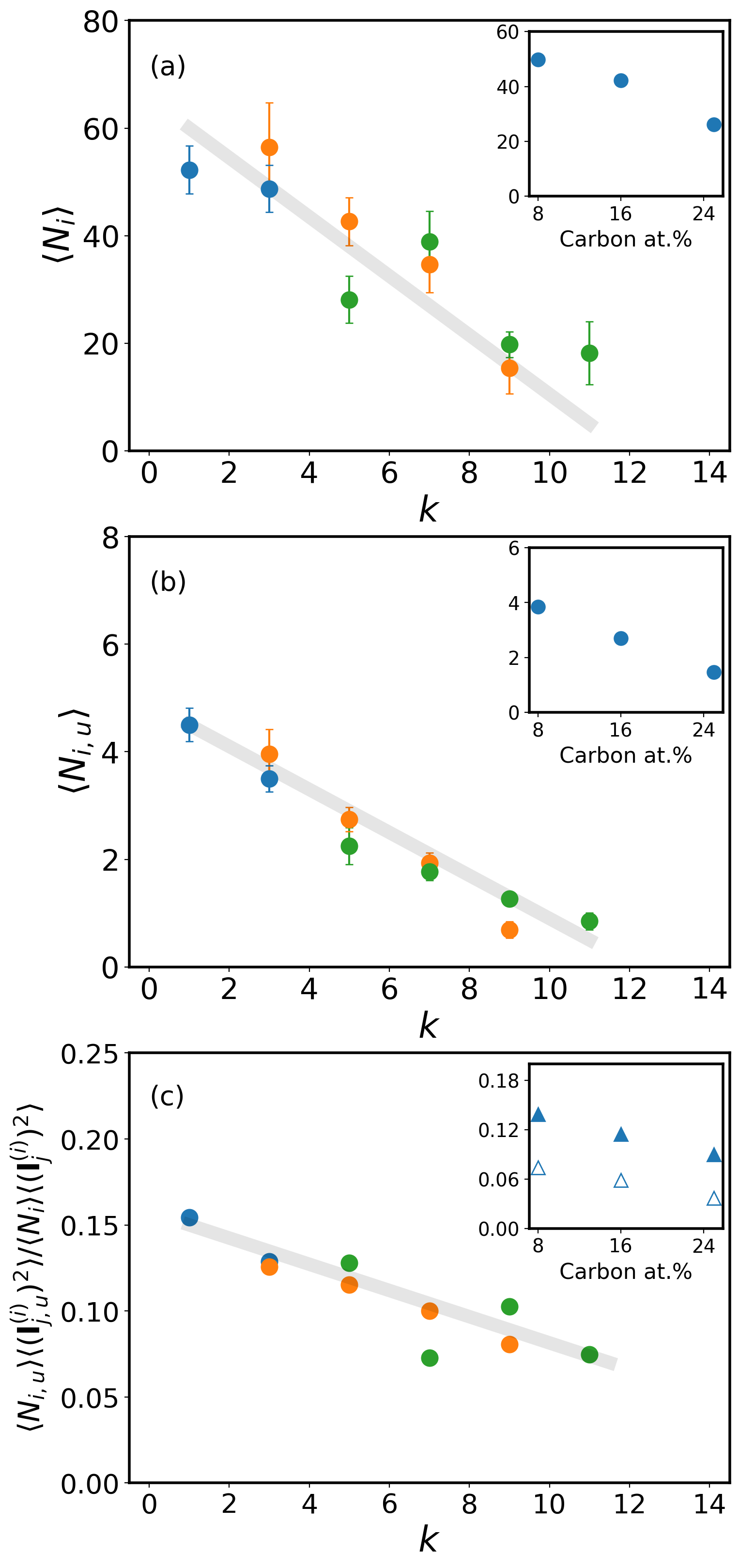}
\caption{(a) The average number of carbon atom hops detected after 200~ns separated into bins based on the number of carbon atoms ($k$) within a neighbour distance of $\SI{4}{\angstrom}$ at the time of the jump of a given atom. (b) Same as (a) but after removing highly correlated jumps. The insets in (a) and (b) show the total average number of jumps observed in 200~ns for each glass. (c) the ratio of the data in (a) and (b), this being related to the correlation factor $f$, also separated into bins based on the number of neighbour carbon atoms at the time of the observed jump.  The inset in this case shows the ratio in the case of the total average number of jumps per atom (open symbols).  The values of the correlation factors for the three glasses determined based on equation~(\ref{eq:timecorrelation}) at 200~ns are shown as filled symbols in the same inset. Lines are shown only as a guide to the eye. Blue corresponds to data from the 8 at.\%C glass, orange to the 16at.\%C glass and green to 25 at.\%C glass. In all cases the data are shown for a threshold of $\cos{\left(\theta\right)} > -0.8$.  }
\label{fig:avejump}
\end{figure}

Figures \ref{fig:avejump}a and b show that both total and uncorrelated number of hops per atom decrease strongly as the number of carbon atoms in their immediate environment increases. Given that the total number of hops is much larger than the number of uncorrelated hops, the total number of hops is nearly identical to the number of correlated hops detected.  In both cases the trend appears to be independent of the bulk carbon concentration of the glass as a single straight line approximates the results obtained from the three glasses.  This would suggest that the composition dependence of the MSD (figure \ref{fig:MSD}) is not due to intrinsic differences in the structure of the glasses tied to the bulk composition, but rather to the local number of carbon neighbours (figure \ref{fig:poisson}).

A linearly decreasing dependence of average number of (uncorrelated) hops with $k$ would naturally arise in a system where site-blocking is prevalent. In the Fe-C system, the strong repulsion between carbon atoms at short separations \cite{GlassyMetalsII,sheng2012Relating} means that trajectories bringing a carbon atom close to another would be highly unfavourable. Under these circumstances, the probability of a hop trajectory being 'unsuccessful' due to its proximity to another carbon atom will be proportional to the number of carbon atoms within the immediate vicinity of the hop.  

This explanation would justify the dependence of the number of observed uncorrelated jumps on the local, and global, carbon composition of the glass. It does not, however, explain the carbon concentration dependence of the correlation factor, $f$. The results in figures \ref{fig:avejump}a and b give another way to estimate $f$. Equating equation~(\ref{eqn:uncorr_factorized}) and equation~(\ref{eqn:corrfac_factorized}), accepting that these are only approximations to the true MSD (cf. figures \ref{fig:MSD_Nhops}a and b) we obtain,

\begin{equation}
    f(t) \approx \langle N_{i}(t)\rangle \langle ({\bf l}^{(i)}_j)^2 \rangle / \langle N_{i,u}(t)\rangle \langle ({\bf l}^{(i)}_{j,u})^2 \rangle 
    \label{corrfac:alt}
\end{equation}
Using this and taking the ratio of the data in figures \ref{fig:avejump}a and b, provides an estimate of $f$, evaluated at 200~ns, as a function of $k$ (figure \ref{fig:avejump}c).  

\begin{figure}[htbp]
\centering
\includegraphics[width=0.9\textwidth]{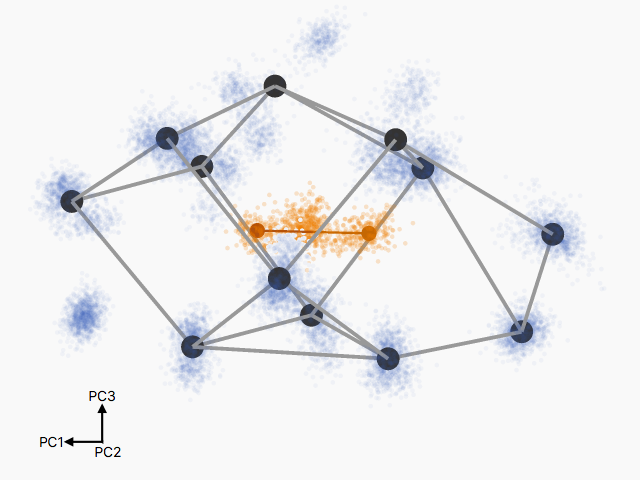}
\caption{An example of a carbon atom observed every 2~ps for 2~ns surrounded by neighbouring Fe atoms.  The large, dark blue spheres show the average location of the Fe atoms, while the diffuse cloud of small blue points shows the actual Fe positions observed every 2~ps.  Similarly, the diffuse cloud of orange points indicates the position of the carbon atom observed every 2~ps.  A large number of long ($> 1.3 $ \AA{}) jumps of the carbon atom were observed, these being highly correlated.  The average start and end positions of these correlated jumps are shown as the large orange points connected by the line.  The distance between these points is $\SI{1.46}{\angstrom}$.  The atom positions have been positioned such that the horizontal direction is the direction along which the jump distance is maximum.   }
\label{fig:correlatedhops}
\end{figure}

First, it is notable that back extrapolating to $k=0$ indicates a very low ($\sim$ 0.17) correlation factor even in the absence of nearby carbon.  Thus, the origins of this large propensity for correlated hops is not intrinsically related to the presence of other carbon atoms. The small size of the carbon relative to the nearest neighbour Fe-Fe (and Fe-C, C-C) distances implies that many carbon atoms are surrounded by significant amounts of space.  This `void space' and its connectivity within glasses and liquids has been linked to the structural ordering of glasses \cite{sheng2006atomic}. Within this connected space between Fe atoms, carbon atoms are able to make hops between adjacent sites,  these being preferentially aligned along directions which are free of closely neighbouring atoms.  An example is shown in figure \ref{fig:correlatedhops}, taken from the glass containing 25 at.\%C, where the position of carbon (orange) and iron (blue) atoms observed every 2~ps over a period of 2~ns are shown.  For this particular carbon atom, it was found that for jumps longer than 1.3 \AA{}, 62\% were found to by highly correlated ($\theta_{i,i+1} > 150^\circ$).  The average location of the end points of these correlated jumps is shown by the large orange points in figure \ref{fig:avejump}.

The dependence of the estimated correlation factor on local carbon concentration (figure \ref{fig:avejump}c) shows the same, bulk composition independent, decay with $k$ as shown in  figure \ref{fig:avejump}a and b. Site blocking, cited above as a possible explanation for the composition dependence of the number of uncorrelated jumps, alone would not be expected to create correlations between jumps. In this case, the explanation may relate to the observed dependence of the local ordering around carbon atoms with $k$, figure \ref{fig:voro}.  Recalling figure \ref{fig:voro}, it was shown that the presence of the dominant tri-capped trigonal prisms surrounding carbon atoms (Voronoi indices $\left<0,3,6,0\right>$) also decreased linearly with $k$, again independent of the bulk composition of the glass. Sheng \emph{et al.} \cite{sheng2012Relating} examined the relationship between `void space' and structural ordering in simulated Ni-P glasses.  It was found in these studies that increasing structural order leads to large voids being reduced in size or annihilated. On top of this, changes in the structural ordering impact on medium range ordering; the way in which the structural polyhedra fill space.  In the case of Fe-B glasses it was shown that the addition of B led to decreasing icosahedral ordering of Fe due to the need to preserve space filling \cite{ganesh2008ab}. Such structural changes to the glass would be expected to impact on the amount of connected void space available for correlated jumps, but the impact of such changes on the average number of jumps is difficult to discern.  

\section{Summary}

By means of molecular dynamics simulations and statistical analysis of individual atomic hops, we have shown a strong composition dependence of the atomic mobility of carbon in an Fe-C glass, this being qualitatively in agreement with experimental observations. The origins of this composition dependence were traced to two sources; the decreasing number of productive (uncorrelated) atomic jumps and an increasing degree of correlation with atomic hops.  The composition dependence was also shown to be related to the local composition rather than the bulk composition of the glass; carbon atoms having a large number of neighbouring carbon atoms were much less likely to make atomic hops than those with no neighbouring carbon atoms.  It has been suggested that the decrease in total number of productive (uncorrelated) jumps could result from site blocking, carbon atoms having strong repulsion at close proximity. The number of correlated jumps, on the other hand, has been hypothesized to be more strongly tied to changes in the underlying structure of the glass.

To return to the central question posed at the beginning of this work, the diffusion of carbon in a model Fe-C glass has been shown here to be strongly carbon concentration dependent, the mobility of the solute atoms decreasing with increasing bulk carbon concentration.  This would appear to be coherent with experimental observations which show a continuous slowing of $\alpha$-Fe growth during the devitrifaction of a binary Fe-C glass. Our results would suggest that the slowing growth is attributable to the accumulation of carbon, rejected from the crystalline $\alpha$-Fe, at the glass/crystal interface.  As carbon accumulates at the interface, mobility of carbon atoms  also reduces leading eventually to an apparent halt to the crystallization. This interpretation would be consistent with the experimental observation of a sudden jump in growth rate attending an increase in annealing temperature. This strong coupling between composition dependence of solute diffusion in the glass and the interface migration rate during devitrification is a point that has largely been ignored in attempts to develop models to predict rates of crystallization.

\section{Acknowledgements}
The authors acknowledge funding for this work from the Natural Science and Engineering Research Council of Canada. JR thanks the Alexander von Humboldt Foundation for financial support.
\section{Bibliography}

\bibliographystyle{unsrt}
\bibliography{references}
\end{document}